\documentclass[twocolumn, tighten, longauthor]{myaastex62}
\usepackage{amsmath,amsfonts,amssymb,amsthm}
\shorttitle{The colors of optical SETI}
\shortauthors{Michael Hippke}
\begin{document}
\title{INTERSTELLAR COMMUNICATION. X. THE COLORS OF OPTICAL SETI}
\author[0000-0002-0794-6339]{Michael Hippke}
\affiliation{Sonneberg Observatory, Sternwartestr. 32, 96515 Sonneberg, Germany}
\email{michael@hippke.org}

\begin{abstract}
It has recently been argued from a laser engineering point of view that there are only a few magic colors for optical SETI. These are primarily the Nd:YAG line at $1{,}064\,$nm and its second harmonic (532.1\,nm). Next best choices would be the sum frequency and/or second harmonic generation of Nd:YAG and Nd:YLF laser lines, 393.8 nm (near Fraunhofer CaK), 656.5\,nm (H$\alpha$) and 589.1\,nm (NaD2). In this paper, we examine the interstellar extinction, atmospheric transparency and scintillation, as well as noise conditions for these laser lines. For strong signals, we find that optical wavelengths are optimal for distances $d\lesssim\,$kpc. Nd:YAG at $\lambda=1{,}064\,$nm is a similarly good choice, within a factor of two, under most conditions and out to $d\lesssim3\,$kpc. For weaker transmitters, where the signal-to-noise ratio with respect to the blended host star is relevant, the optimal wavelength depends on the background source, such as the stellar type. Fraunhofer spectral lines, while providing lower stellar background noise, are irrelevant in most use cases, as they are overpowered by other factors. Laser-pushed spaceflight concepts, such as ``Breakthrough Starshot'', would produce brighter and tighter beams than ever assumed for OSETI. Such beamers would appear as naked eye stars out to kpc distances. If laser physics has already matured and converged on the most efficient technology, the laser line of choice for a given scenario (e.g., Nd:YAG for strong signals) can be observed with a narrow filter to dramatically reduce background noise, allowing for large field-of-view observations in fast surveys.\\
\end{abstract}

\section{Introduction}
Only months after the seminal paper on radio SETI \citep{1959Natur.184..844C}, the laser was discovered by \citet{1960Natur.187..493M}, and interstellar optical communication was proposed \citep{1961Natur.190..205S}. While lasers are too energetically costly as isotropic or wide-angle beacons, their tighter beams are useful for directed communications \citep{2017arXiv171105761H}.

The justification for the feasibility of optical SETI is typically given with the example that current lasers and telescopes would be sufficient to signal over large distances \citep[e.g.,][]{2004ApJ...613.1270H}. Indeed, a laser focused and received through 10\,m telescopes would deliver one photon per kJ pulse at $\lambda=500\,$nm over a distance of 100\,pc. The strongest lasers on Earth supply $\sim2\,$MJ in a 5\,ns ($5\times10^{-9}\,$s) pulse, although at low (hrs) repetition rates \citep{2017ApPhB.123...42H}. A Sun-like G2V star at a distance of 100\,pc delivers $2\times10^8$ photons per second into a 10\,m telescope on Earth, less than one per ns observation cadence, so that a MJ laser outshines the host star by a factor of $10^4$ during a pulse, independent of distance.

It has recently been argued, from a laser engineering point of view, that the optimal transmitter is a solid-state high power laser such as Neodymium Yttrium Aluminum Garnet (Nd:YAG) or Neodymium Yttrium Lithium Fluoride (Nd:YLF) \citep{2018NewA...60...61N}. These can be operated in both pulsed and continuous mode. Although there are methods which allow to tune laser light over a very large wavelength band, such as the free electron laser, this is energetically inefficient and expensive \citep{2010IJQE...46.1135S,2016PhRvS..19b0705E,2016PhRvL.117q4801S}. If the optimal fixed wavelength can not be used, e.g. due to atmospheric absorption given a certain (exo)planet chemistry, the next best choices have been argued to be the second harmonic generation and/or sum frequency generation of YAG and YLF lasers \citep[][our Table~\ref{tab:laser_lines}]{2018NewA...60...61N}. A laser in the H$\alpha$ spectral line (656.2808\,nm) was already suggested by \citet{1993SPIE.1867...75K} and \citet{Ross1993}. Using laser frequencies which coincide with spectral absorption lines has been argued to be more \citep{asimov1979extraterrestrial} or less \citep{1993SPIE.1867...75K} useful, but without any quantitative assessment.

It might be the case that human laser engineering has reached a maturity \citep{Koechner2006} from which we can deduce the ``optimal'' laser lines in terms of signal power. We can compare the throughput of photons between transmitter and receiver for a specific laser line, limited by extinction, atmospheric transparency and noise, with the available transmitter power. If a certain laser line offers e.g., $3\times$ the output power for a finite monetary investment, but suffers from $2\times$ higher extinction, it would still be preferable as it maximizes the number of photons at the receiver for a given budget. To our knowledge, individual laser lines have never been studied in detail with respect to OSETI. This is the aim of the paper: To determine the best wavelength in order to optimize future surveys.\\

\begin{table*}
\center
\caption{``Magic wavelengths'' in OSETI\label{tab:laser_lines} with losses.}
\begin{tabular}{clccccr}
\tableline
$\lambda$ (nm) & Laser type & $S_{\rm E}$ (100\,pc)& $S_{\rm E}$ (kpc) & $S_{\rm A}$ & $F_{\rm r} $& Comment  \\
\tableline
1064.1  & YAG (R$_2\rightarrow$Y$_3$)      & 0.94 & 0.54 & 0.96 & 0.37 & Most common YAG \\
532.1   & SHG--YAG (R$_2\rightarrow$Y$_3$) & 0.84 & 0.18 & 0.83 & 0.74 & Second harmonic \\
393.8   & SFG[YAG(R$_2\rightarrow$X$_3$)+SHG-YAG(R$_1\rightarrow$Y$_5$)] & 0.77 & 0.08  & 0.63 & 1.00 & Close to CaK line \\
656.5   & SHG--YLF$(\sigma)$ 1.313$\,\mu$m & 0.88 & 0.27 & 0.90 & 0.60 & Close to H$\alpha$ line \\
589.1   & SFG[YAG(R$_1\rightarrow$Y$_2$) + YAG(R$_2\rightarrow$X$_1$)]& 0.86 & 0.22  & 0.85 & 0.67 & Close to NaD2 line \\
\tableline
\end{tabular}
\\Sorted in descending order of likelihood, following \citet{2018NewA...60...61N}. $S_{\rm E}$ is the fraction of photons which defies interstellar extinction (section~\ref{sub:ext}). $S_{\rm A}$ is the fraction which defies atmospheric absorption, here for optimal conditions (section~\ref{sub:atmo}). $F_{\rm r}$ is the fraction which defies extinction, compared to $F_{\rm r, \lambda=393.8\,nm}\equiv1$ (section~\ref{sub:beam}). 
\end{table*}

\begin{figure*}
\includegraphics[width=.5\linewidth]{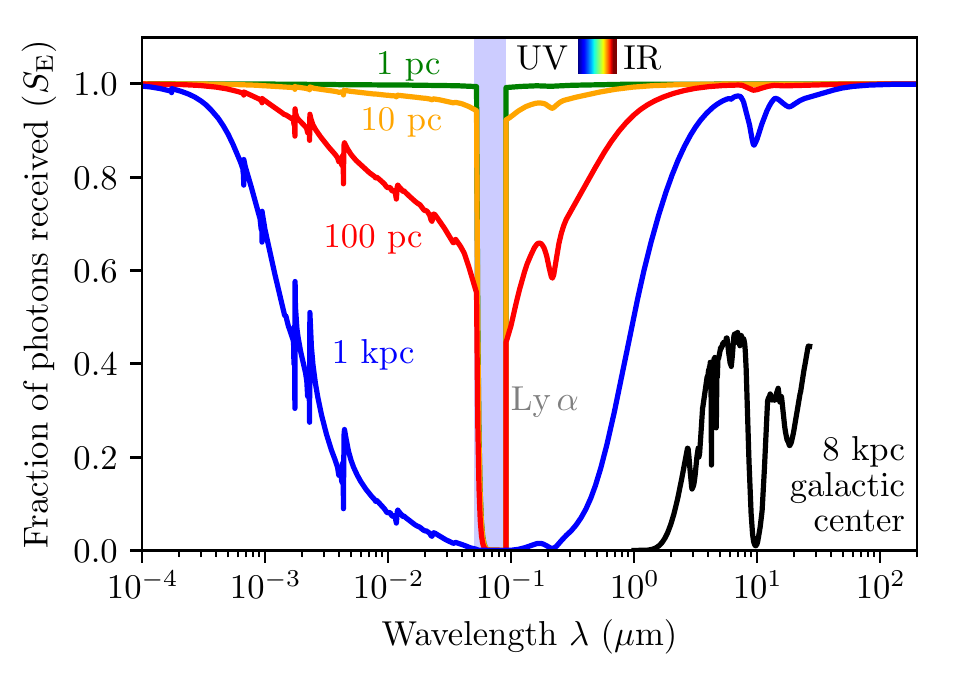}
\includegraphics[width=.5\linewidth]{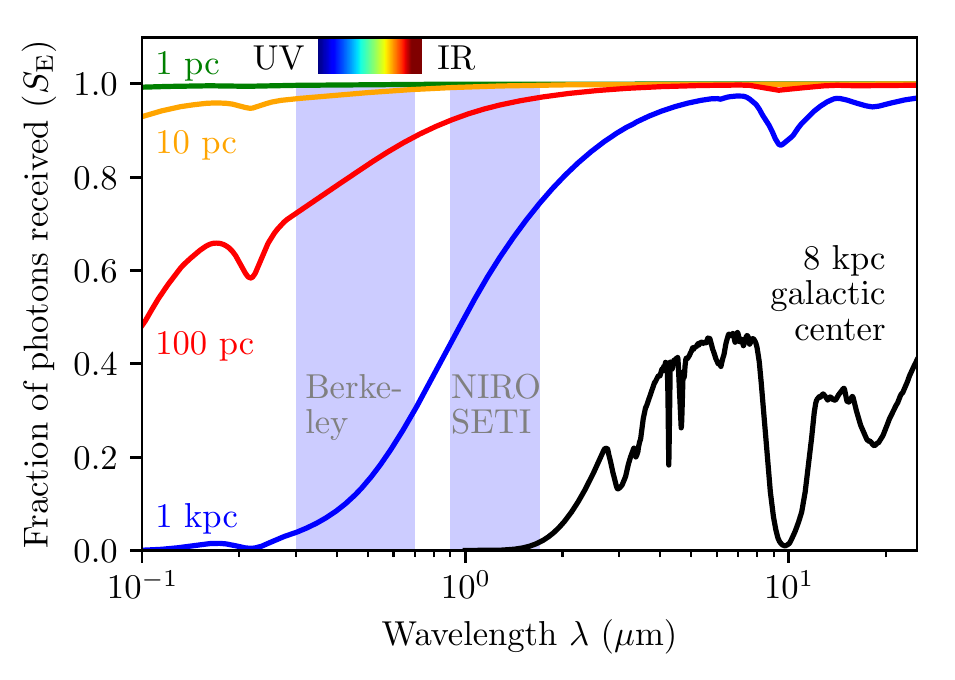}
\caption{\label{fig:oseti_extinction}Interstellar extinction. Earth's atmosphere is opaque for wavelengths short of UV (left half in left panel). Right panel: Optical to UV region with two major OSETI programs highlighted.\\}
\end{figure*}

\section{Losses}
We will now trace the loss of photons along their path from the transmitter through interstellar space, Earth's atmosphere, and into a detector.

\subsection{Beam diffraction}
\label{sub:beam}
An electromagnetic beam widens with distance due to diffraction. A receiver with aperture $D_{\rm r}$ obtains a flux of \citep{kaushal2017free,2017arXiv170603795H}

\begin{equation}
\label{eq2}
F_{\rm r} = \frac{P_{\rm t} D_{\rm t}^2 D_{\rm r}^2}{4 h f Q^2 \lambda^2 d^2} ({\rm s}^{-1})
\end{equation}

over a free-space distance $d$, where $P_{\rm t}$ is the transmitted power, $f$ the photon frequency, and $h$ Planck's constant. The (half) opening angle of the diverging light beam is $\theta = Q \lambda/D_{\rm t}$ (in radians) with $Q\sim1.22$ for a diffraction limited circular transmitting telescope of diameter $D_{\rm t}$ \citep{rayleigh}, and $\lambda=c/f$ with $c$ as the speed of light.

Photon energy depends on wavelength, $E=hc/ \lambda$, which makes higher frequency photons more costly in terms of energy. The beam angle, however, decreases linearly for higher frequencies, and as a consequence the flux decreases quadratically for the area in the beam. For a given aperture size, distance and sufficiently smooth surface, the received number of photons scales as $\gamma \propto f$ so that higher frequencies (shorter wavelengths) have a positive linear relation to the number of photons in the receiver \citep{2017arXiv171105761H}.

\subsection{Interstellar extinction}
\label{sub:ext}
At UV, optical and IR wavelengths, extinction is due to scattering of radiation by dust. Wavelengths shorter than the Lyman limit ($91.2\,$nm) are affected by the photo-ionisation of atoms \citep{1996Ap&SS.236..285R}. Extinction levels depend on distance and galactic latitude, with average values of $\sim0.1$\,mag over 100\,pc in the optical in the solar neighborhood, which increases to $0.05\dots0.15$\,mag at 200\,pc \citep{1998A&A...340..543V}.

We show synthetic extinction curves in Figure~\ref{fig:oseti_extinction} using data from  \citet{2003ARA&A..41..241D,2003ApJ...598.1017D,2003ApJ...598.1026D,2004ApJ...616..912V,2011ApJ...737...73F,2016ApJ...828...69M} as discussed in \citet{2017arXiv170603795H}. While extinction is typically given in astronomical magnitudes, we convert these to the fraction of photons received over distance ($S_{\rm E}$). Quantitatively, extinction becomes relevant ($S_{\rm E}<0.5$) for distances of $\gtrsim300$\,pc in the optical and $\gtrsim$\,kpc in the NIR. OSETI in the visual is only sensible for short ($\lesssim$\,kpc) interstellar distances. For longer distances, near-infrared searches are preferred \citep[``NIROSETI'',][]{2014SPIE.9147E..0JW,2014SPIE.9147E..4KM,2016SPIE.9908E..10M}.

Towards the galactic center, extinction increases to large values, $E(B-V)\approx3$ at $A(V)>44$\,mag at 550\,nm \citep{2008A&A...488..549P,2011ApJ...737...73F}, an attenuation by a factor of $10^{-18}$. Towards the center of the galaxy, mid-IR wavelengths ($\sim5-8\,\mu$m) are preferable for communication, while the absorption lines of water ice ($3.1\,\mu$m) and silicate ($10\,\mu$m, $18\,\mu$m) should be avoided (Figure~\ref{fig:oseti_extinction}, right panel).

\begin{figure*}
\includegraphics[width=\linewidth]{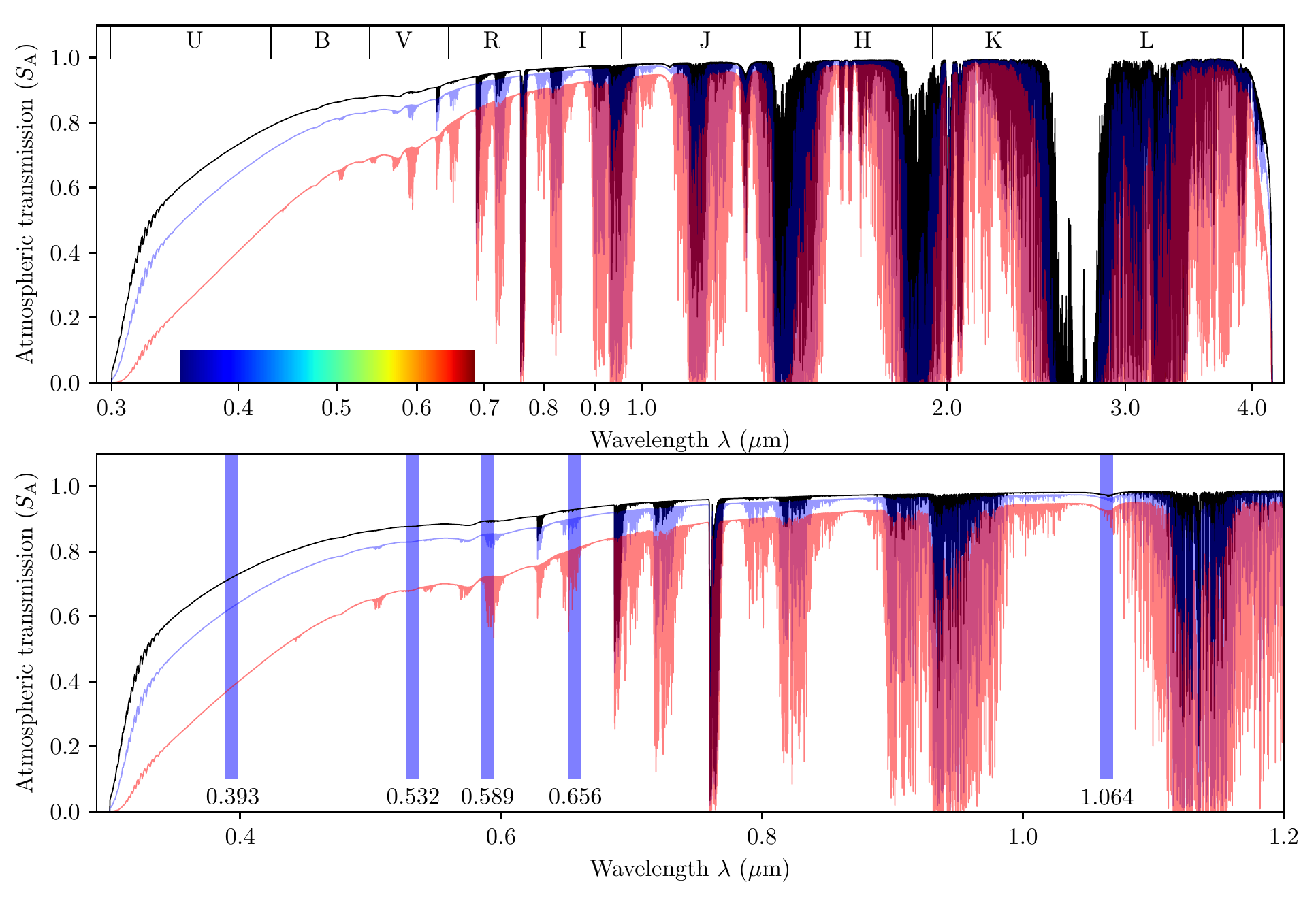}
\caption{\label{fig:oseti_atmo1}Transmission through Earth's atmosphere as a function of wavelength at Cerro Paranal (VLT site) for different conditions. Best possible conditions in black with 0.5\,mm precipitable water vapor and zenith angle. Blue shows median conditions (2.5\,mm), while red has the lower quartile conditions (5\,mm) at $20^{\circ}$ zenith angle. Top: Optical to mid-IR. Bottom: Zoom into optical to near-IR, with the laser lines in question shown as blue vertical bars.}
\end{figure*}

\begin{figure*}
\includegraphics[width=.5\linewidth]{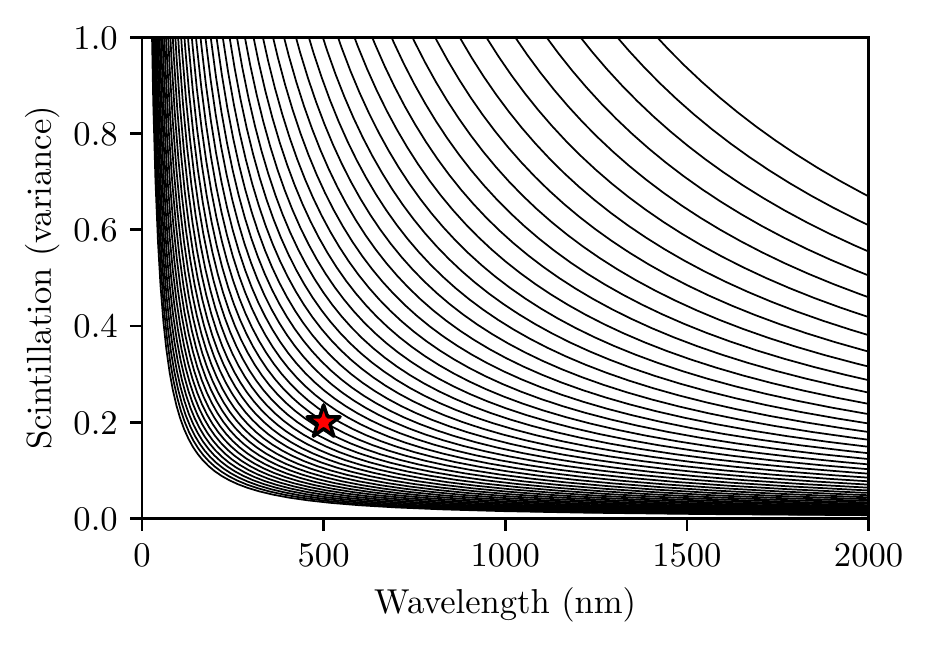}
\includegraphics[width=.5\linewidth]{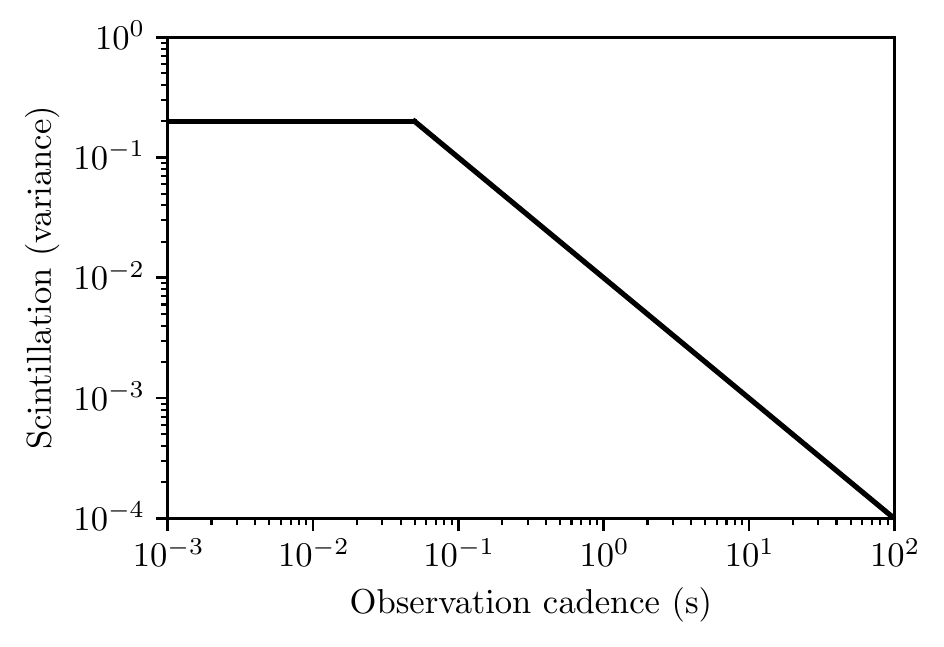}
\caption{\label{fig:scinti}Left: Estimated scintillation as a function of wavelength, with curves for medium to low turbulence ($10^{-15}<C^2<10^{-17}$) from space to ground. The red symbol shows the measurements taken with the 1\,m Jacobus Kapteyn Telescope on La Palma under typical conditions \citep{2015MNRAS.452.1707O}. Right: Scintillation as a function of observational cadence, scaled following Figure 10 in \citet{2015MNRAS.452.1707O}.}
\end{figure*}

\subsection{Atmospheric transparency}
\label{sub:atmo}
Realistic laser beams are so narrow that they are targeted at one planet at a time. For example, an aperture $D_{\rm t}=10\,$m transmitting with a wavelength $\lambda=\,\mu$m produces a beam with $\theta=25\,$mas. The cone broadens to 1\,au at a distance $d\sim40\,$pc. Targeting one specific planet allows to optimize the transmitted wavelength to its unique atmospheric transparency and noise features. Atmospheric transmission depends on the wavelength and varying characteristics, such as the content of water vapor in the air, which mainly depends on the altitude. Space telescopes are ideal, but much more expensive than ground-based telescopes, and offer little advantage for OSETI; a fact ET knows and supposedly takes into account for its messaging.

ET can infer Earth's atmospheric composition, and thus transparency, from high resolution spectroscopy, assuming Earth and Sun can be spatially resolved. If Earth is a transiting planet as seen from ETs observing point of view \citep{1988ATsir1531...31F,1990ATsir1544...37F,2016AsBio..16..259H}, transit spectroscopy allows to determine Earth's atmosphere. Such observations are currently being investigated for nearby exoplanets such as Proxima Cen, and are within reach for next-generation telescopes and instruments \citep{2017A&A...599A..16L}. Therefore, we can assume that with superior technology, such observations are possible over large (e.g., kpc) distances. What ET however can not infer are the political and economical locations of telescopes. Atmospheric transparency is a strong function of wavelength and water vapor, with altitude being beneficial. Mountains might be detectable through transit observations \citep{2018MNRAS.tmp..142M}. It is however remotely unknown whether telescopes can be placed on Earth's mountain tops, or whether lower altitude locations are required for political, economical, or other reasons. Conservatively, ET should not assume best possible atmospheric conditions, but rather optimize towards a wavelength which is also received on most of the surface.

To estimate atmospheric transparency, we use SkyCalc\footnote{\url{http://www.eso.org/observing/etc/skycalc/skycalc.htm}}
based on the Cerro Paranal Advanced Sky Model \citep{2012A&A...543A..92N,2013A&A...560A..91J}. It is based on theoretical models, calibrated with real observational data, and available at a resolution of $R=10^6$ for $0.3<\lambda<30\,\mu$m.

Atmospheric transmission is essentially zero for wavelengths $\lesssim300\,$\,nm, above 20\,m, and between $30\dots200$\,$\mu$m. In the optical, it increases steadily between 300\,nm and 650\,nm. In the infrared, transparency fluctuates rapidly between near unity and zero due to numerous absorption lines from water, carbon dioxide, ozone and other gases (Figure~\ref{fig:oseti_atmo1}). Communications with a narrow (nm) bandwidth require a careful choice of the wavelength. For example, a high-power Tm:YAG diode-side-pumped rod laser at $2.07\,\mu$m might be a power-efficient choice \citep{Wang2013}, however atmospheric transmission is only $\sim20\,$\% ($\sim10\,$\%) in best (median) conditions at Cerro Paranal. A similar Ho:YAG laser at $2.08\,\mu$m results in $>90\,$\% transparency under all conditions. Clearly, one would not expect any signals at unfavorable wavelengths, and one might consider blocking these bands with filters to reduce background radiance.

As can be seen in Figure~\ref{fig:oseti_atmo1}, observing conditions are mostly relevant at the short wavelength end ($\lambda=393.8\,$nm). With low atmospheric water vapor content, transparency can be $\gtrsim70\,$\%, but deteriorates by a factor of $\sim2$ for median conditions to $\lesssim35\,$\%. Differences in atmospheric quality become negligible for longer wavelengths (excluding the IR absorption bands), and are $\lesssim 5\,$\% at $\lambda=1\,\mu$m.

If the laser bandwidth is small (nm), many good IR choices exist with transparency $>90\,$\%. The broadest such feature is between $0.99 \dots 1.09\,\mu$m, so that a $\Delta \lambda\lesssim50\,$nm Nd:YAG laser centered at 1064\,nm is an excellent choice. In the IR, there are bands with high transparency around $750\pm5, 780\pm5, 865\pm20, 1240\pm5, 1280\pm5, 1550\pm10, 1590\pm3$\,nm.

Atmospheres of other inhabited exoplanets likely differ from Earth's atmospheric composition, producing a different transparency as a function of wavelength. It would be interesting to compare the preferred transparency windows of other habitable worlds to check if the Nd:YAG laser line is universally suitable.

\subsection{Atmospheric scintillation}
Stellar scintillation (``twinkling'', ``flickering''), is the apparent brightness (and position) variation viewed through the atmospheric medium. It is caused by anomalous refraction through small-scale fluctuations in air density due to temperature gradients. The strength of scintillation is typically measured in terms of the variance of the beam amplitude \citep[or irradiance, Rytov variance,][]{1988ApOpt..27.2150A}

\begin{equation}
\sigma^2 = 1.23C^2 k^{7/6} H^{11/6}
\end{equation}

where $k = 2 \pi / \lambda$ is the wave number, H is the scale height of the atmospheric turbulence, generally accepted to be $H\sim8{,}000\,$m \citep{2015MNRAS.452.1707O}, and $C^2$ is the structure constant for refractive-index fluctuations as a measure of the optical turbulence strength.
Measured values are $C^2=1.7\times10^{-14}\,{\rm m}^{-2/3}$ at ground level and $C^2=2\times10^{-18}\,{\rm m}^{-2/3}$ at a height of 14\,km \citep{Coulman1988}. A vertical profile in the Negev desert between ground level and 20\,km altitude has values between $10^{-16}<C^2<10^{-15}$ \citep{Kopeika2001,Zilberman2001}. Scatter in measurements from different locations is larger than the height dependence. Often quoted distinctions for the turbulence are $10^{-13}$ (strong),  $10^{-15}$ (average), and $10^{-17}$ (weak) \citep{goodman1985statistical,XiaomingZhu2002}. Turbulence is particularly low at Dome C in Antarctica, about $2 \dots 4 \times$ lower than at Cerro Tololo and Cerro Pach\'{o}n in Chile \citep{2006PASP..118..924K}, while other major observatory sites such as La Palma, Mauna Kea, Paranal and Tololo are within a factor of two of each other \citep{2015MNRAS.452.1707O}.

Longer wavelengths experience a smaller variance. Figure~\ref{fig:scinti} shows contours for scintillation variance as a function of wavelength for weak to medium turbulence ($10^{-17}<C^2<10^{-15}$). Scintillation is a factor of a few larger at optical compared to NIR wavelengths.

\subsection{Detector efficiency}
An ideal detector would have a close to 100\,\% detection (quantum) efficiency: The probability that a photon is successfully detected every time it hits the detector. There would be zero ``dark counts''. Its reset time after a detection would be instant, so that the interval between two detections is infinitely short. Finally, the uncertainty of the arrival time would be nearly zero. In practice, technical issues prevent the realization of these goals. Still, current and future detectors offer quantum efficiencies of $\approx50\,$\% in the optical and IR at ns cadence.

Several detector technologies are used to measure individual incoming photons. Sensors differ in their timing and amplitude precision, dead time, photon efficiency, cooling requirements, and other factors. Traditionally, OSETI was performed in the optical using PMTs at ns cadence with sensitivities between $300<\lambda<650\,$nm and quantum efficiencies of $\sim20\,$\%. Common PMTs are e.g., Hamamatsu types R1548 and R3896 with market prices of $1{,}000\,$ USD new (50 USD used).
Today, commercial photomultipliers offer bandwidths of $\approx100\,$\%, quantum efficiencies $\approx50\,$\%, dark rates of a few hundred Hz when cooled, reset times of $\approx3\,$ns and timing jitter of $\approx0.1\,$ns \citep{2010NIMPA.618..139A}.

Recent detector advances extend the wavelength coverage to the IR at higher quantum efficiencies, allow for shorter cadences, and offer multipixel detectors.

The highest efficiencies are offered by superconducting nanowires (SSPDs) at $\sim93\,$\% QE in the IR with a timing jitter of 0.15\,ns \citep{2013NaPho...7..210M}. These require cooling to a few K \citep{JinZhang2003}. SSPDs can provide ultrahigh counting rates exceeding 1\,GHz \citep{2008ApPhL..92x1112T}.

Other technologies, such as InGaAs SPADs have less stringent cooling requirements. They offer excellent timing resolution ($<60$\,ps) with short dead time ($\sim10\,$ns).

The latest developments in Microwave Kinetic Inductance Detectors (MKIDs) provide large arrays at maximum count rates of $\sim10^3$ counts/pixel/s. They cover optical and IR wavelengths but provide only $\mu$s timing and require deep (K) cooling \citep{2012RScI...83d4702M,2012OExpr..20.1503M}.

\begin{figure*}
\includegraphics[width=\linewidth]{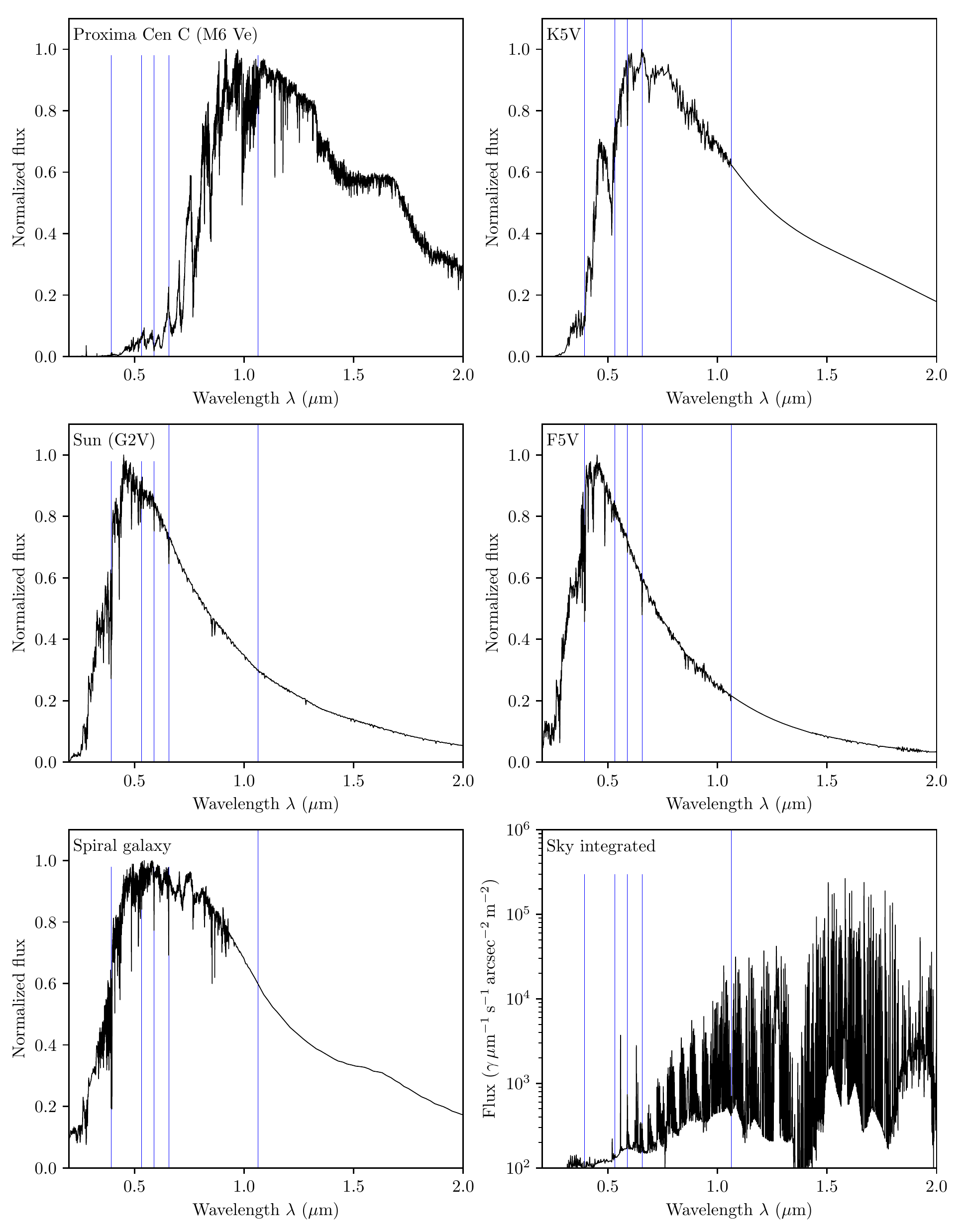}
\caption{\label{figure_oseti_proxima}Noise backgrounds with laser lines shown in blue.
Proxima Cen C from \citet{1998AJ....115..345S,2016arXiv160808620M,2017A&A...603A..58R},
K5V/F5V from \citet{2004ApJS..151..387I}, spiral galaxy from \citet{2013A&A...551A.100B},
solar spectrum from \citet{1981SoPh...74..231N},
sky integrated (zodi and starlight) from \citet{1998A&AS..127....1L,2012A&A...543A..92N,2013A&A...560A..91J}.}
\end{figure*}

\section{Noise}

\subsection{Stellar noise}
The flux from a Sun-like (G2V) isotropic radiator can be approximated in the visual as

\begin{equation}
F \sim 32\times10^{-9} \,{\rm s}^{-1} \,{\rm m}^{-2}
\left(\frac{L}{L_{\odot}}\right)
\left(\frac{d}{1\,{\rm pc}}\right)^{-2}
\end{equation}

where $L$ is the luminosity (in solar luminosities) and $d$ the distance in pc, so that we receive $\sim32$ photons per m$^2$ per nanosecond from a Sun-like star at a distance of one parsec. When observing one star at a time, where the transmitter is typically blended with the host star, this is the astrophysical noise which enters the detector.

Different stellar types have different spectra. For example, M-type stars are the most frequent and have their luminosity peak in the infrared, near $1\,\mu$m, while Sun-like stars have their peak in the optical ($\sim500\,$nm). As can be seen in Figure~\ref{figure_oseti_proxima}, the flux of a M6 dwarf (such as our nearest neighbor, Proxima Cen C) is much less in the optical than in the IR, $F(\lambda=393\,{\rm nm})/F(\lambda=1064\,{\rm nm})\sim0.7\,$\%. This is a significant difference for weaker OSETI transmitters.

Other stellar types are less affected by the spectral energy distribution with respect to the laser lines in question. For Sun-like stars, the difference is $\sim50\,$\%. The Fraunhofer lines (e.g., CaK, H$\alpha$, NaD2) have the advantage of a reduced flux of $\lesssim 50\,$\%.

\subsection{Other galaxies}
The integrated spectral energy distribution of a spiral galaxy, such as Andromeda (Figure~\ref{figure_oseti_proxima}), has a steep flux increase between $200\,$nm and $450\,$nm. There is a broad peak $450\lesssim\lambda\lesssim900\,$nm and spectral lines with depths $\lesssim50\,$\%. The flux decreases slowly towards the IR. Other galaxy types have very different spectra, and would require a dedicated analysis.

\begin{figure*}
\includegraphics[width=\linewidth]{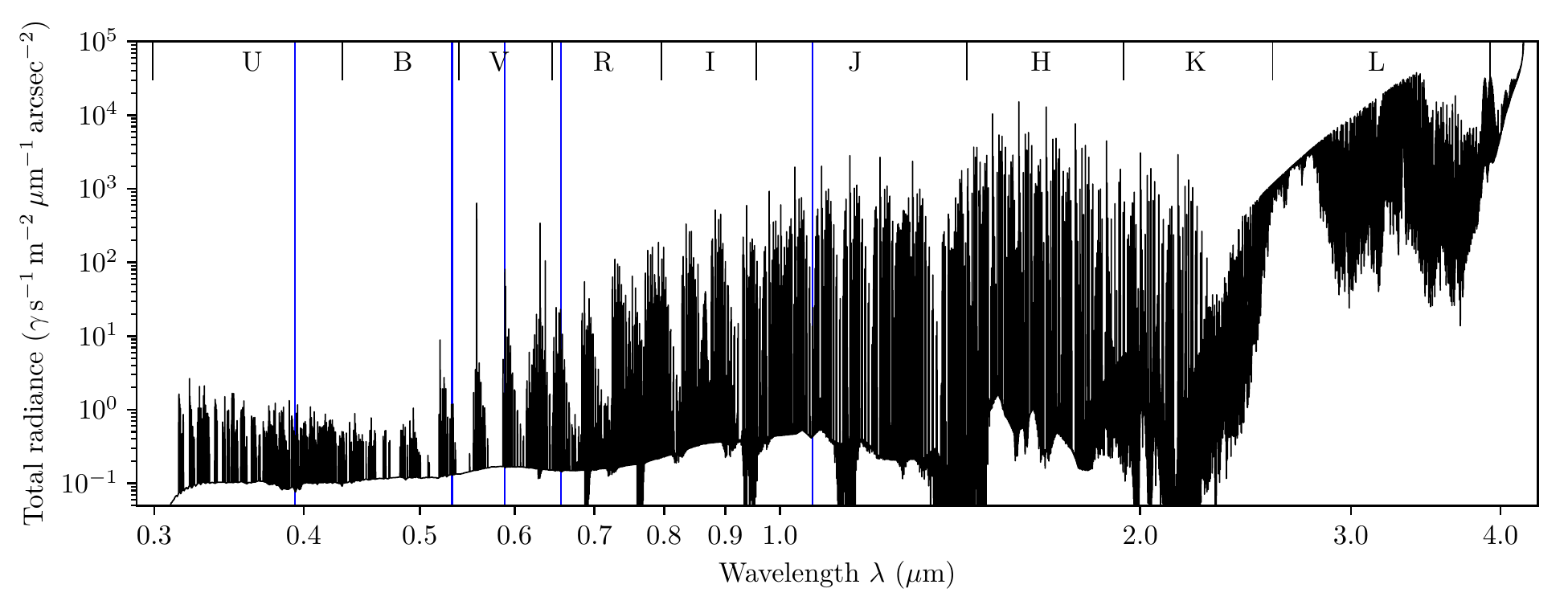}
\caption{\label{figure_oseti_noise_atmo}Atmospheric radiance acting as noise. A zoomed plot is available in Figure~\ref{figure_oseti_noise_atmo_multi}.}
\end{figure*}

\begin{figure}
\includegraphics[width=\linewidth]{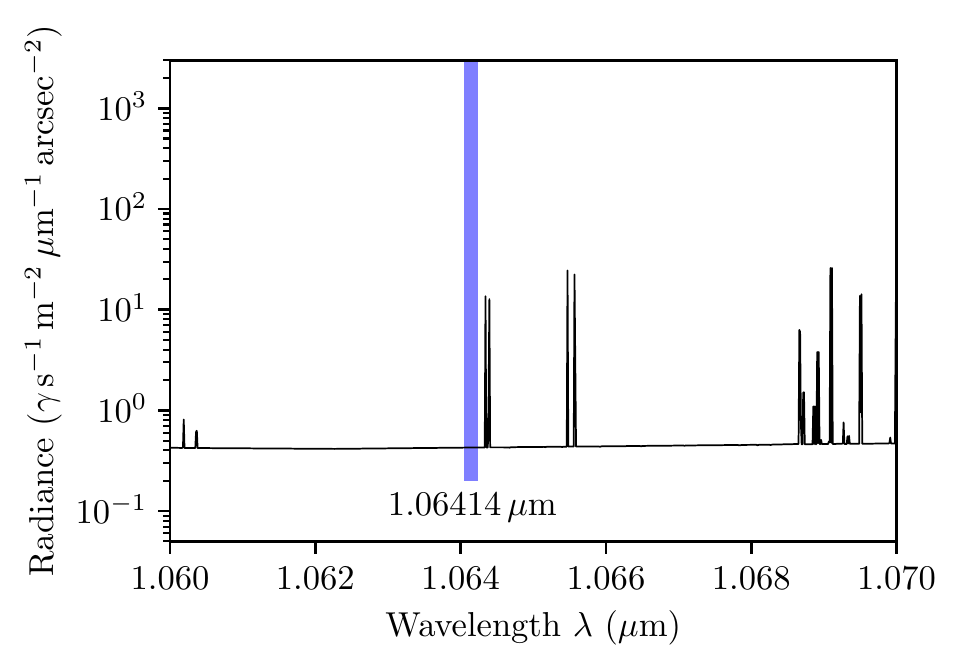}
\caption{\label{figure_oseti_noise_atmo_zoom2}Zoom into atmospheric radiance for the Nd:YAG line with a linewidth of 0.1\,nm. No noise peak coincides with the laser line at this width.}
\end{figure}

\subsection{Atmospheric noise}
\label{sub:atmo_noise}
To estimate the relevant atmospheric noise sources, we again use SkyCalc \citep{2012A&A...543A..92N,2013A&A...560A..91J} which includes noise sources for scattered moonlight, starlight, zodiacal light, molecular emissions from the lower atmosphere, sky emission lines of the upper atmosphere, and the airglow continuum. We exclude moonlight from our analysis as it overpowers all other sources, and can be avoided in $\sim50\,$\% of the observation time. We select the observatory at VLT Cerro Paranal at an altitude of 2640\,m as the location. We chose median observing conditions with a precipitable water vapor of $2.5\,$mm at zenith angle (Figure~\ref{figure_oseti_noise_atmo}).

\begin{figure*}
\includegraphics[width=\linewidth]{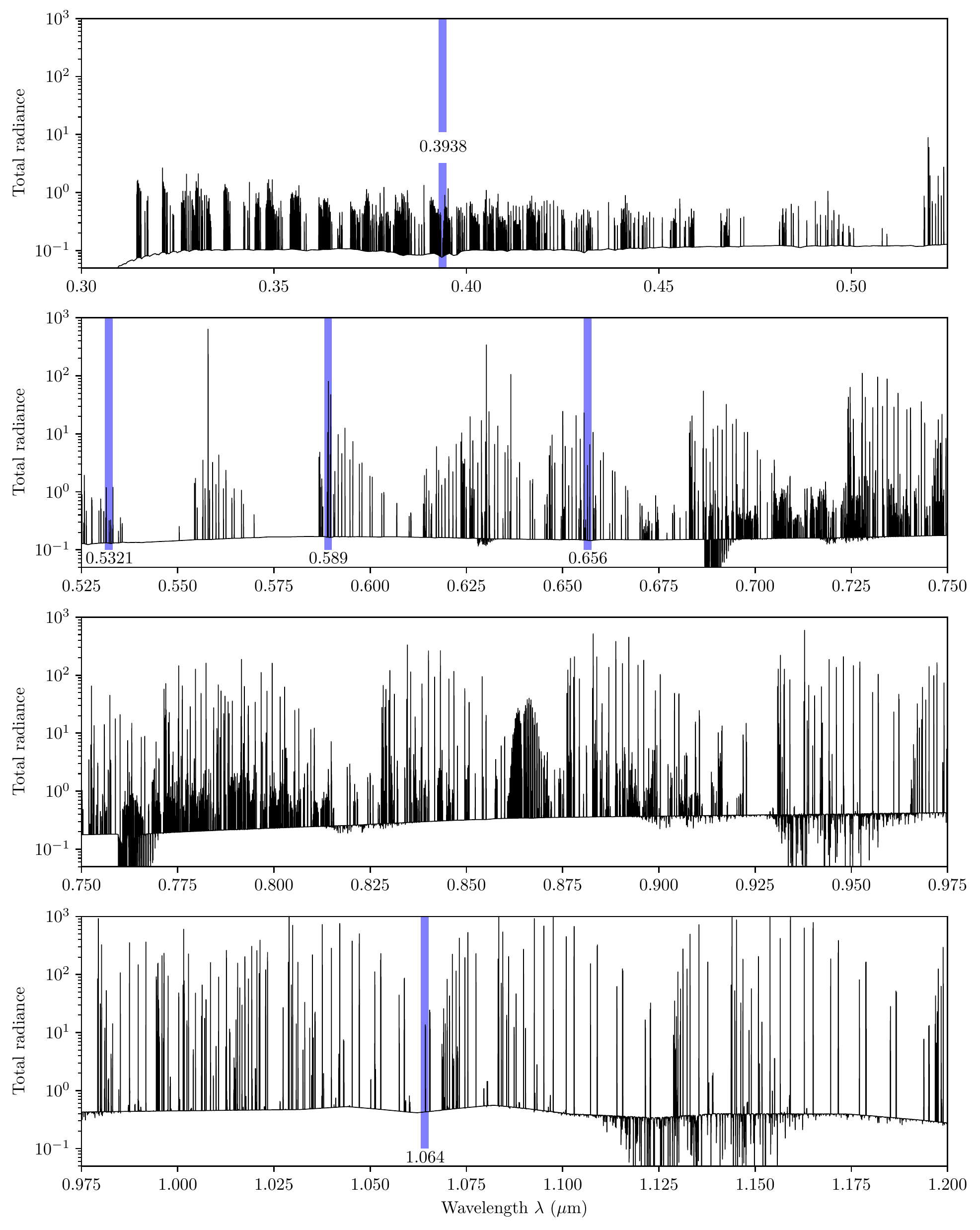}
\caption{\label{figure_oseti_noise_atmo_multi}Zoom into atmospheric radiance (in units of $\gamma\,$s$^{-1}\,$m$^{-2}\,\mu$m$^{-1}\,$arcsec$^{-2}$) as a function of wavelength. The linewidths are 2\,nm.}
\end{figure*}

A zoom into the spectral view shows a forest of narrow noise lines (Figure~\ref{figure_oseti_noise_atmo_multi}). Noise levels fluctuate rapidly by two orders of magnitude over 0.01\,nm wavelength. If the laser lines are broad ($\gtrsim 1\,$nm), these features can be ignored, and an average radiance of $\sim 0.1\dots1\, \gamma\,$s$^{-1}\,$m$^{-2}\,\mu$m$^{-1}\,$arcsec$^{-2}$ can be assumed.

If the OSETI laser lines are narrow ($\lesssim 0.01\,$nm), the wavelength should be chosen by ETI so that they do not occur exactly at a noise peak (Figure~\ref{figure_oseti_noise_atmo_zoom2}). Nd:YAG laser lines can be extremely narrow \citep[$<1\,$Hz,][]{1993OptL...18..505U,Webster2004,Jiang2009}. Then, other factors set an observational limit (e.g., for filters at the receiver), such as Earth's rotation and motion around the Sun ($\Delta \lambda \gtrsim0.1\,$nm), and the time-bandwidth limit from the \citet{Heisenberg1927} uncertainty principle, $\Delta f \Delta t \simeq 1$. A $\Delta t_{\rm min}=1\,$ns pulse has a minimum bandwidth of $\Delta \lambda \gtrsim 10^{-3}\,$nm. Short pulse duration limits will be explored in paper~11 of this series.

\subsection{Atmospheric noise plus sky-integrated starlight}
Instead of observing only one star at a time, a survey could observe a larger field of view \citep{2007AcAau..61...78H,2013PhDT.......161M}. Then, the light of many stars and galaxies is averaged (Figure~\ref{figure_oseti_proxima}, bottom right panel). We use the same model as for the atmospheric noise (section~\ref{sub:atmo_noise}), supplemented with the noise sources of scattered starlight and zodiacal light. Zodi is strongly ($>2$ orders of magnitude) dependent on the heliocentric ecliptic longitude \citep{1980A&A....84..277L,1992ApJ...397..420B,1998ApJ...508...44K}, and we choose ``weak'' settings with a longitude of 135$^{\circ}$ and an ecliptic latitude of 90$^{\circ}$. In the optical, the sky-integrated flux is a few hundred photons $\mu{\rm m}^{-1}\, {\rm s}^{-1}\, {\rm arcsec}^{-2}\,{\rm m}^{-2}$. For a 1\,m telescope observing one entire hemisphere ($2.7\times10^{11}\,$arcsec$^2$) over a bandwidth of $1{,}000\,$nm, the flux is $\sim2\times10^{14}$ photons per s. To make the background noise small at ns cadence ($<10^9\,\gamma\,{\rm s}^{-1}$), one could reduce the sky coverage, bandwidth, or a combination of both. When observing only one laser line, e.g. at $1064.14\,$nm with a bandwidth of 0.1\,nm at ns cadence, the maximum sky coverage can be $\sim3$\,\% of one sky hemisphere. This would allow for fast surveys, scanning the entire visible sky multiple times per night. Another option would be to increase the time resolution, which will be discussed in paper~11 of this series.

\begin{figure*}
\includegraphics[width=\linewidth]{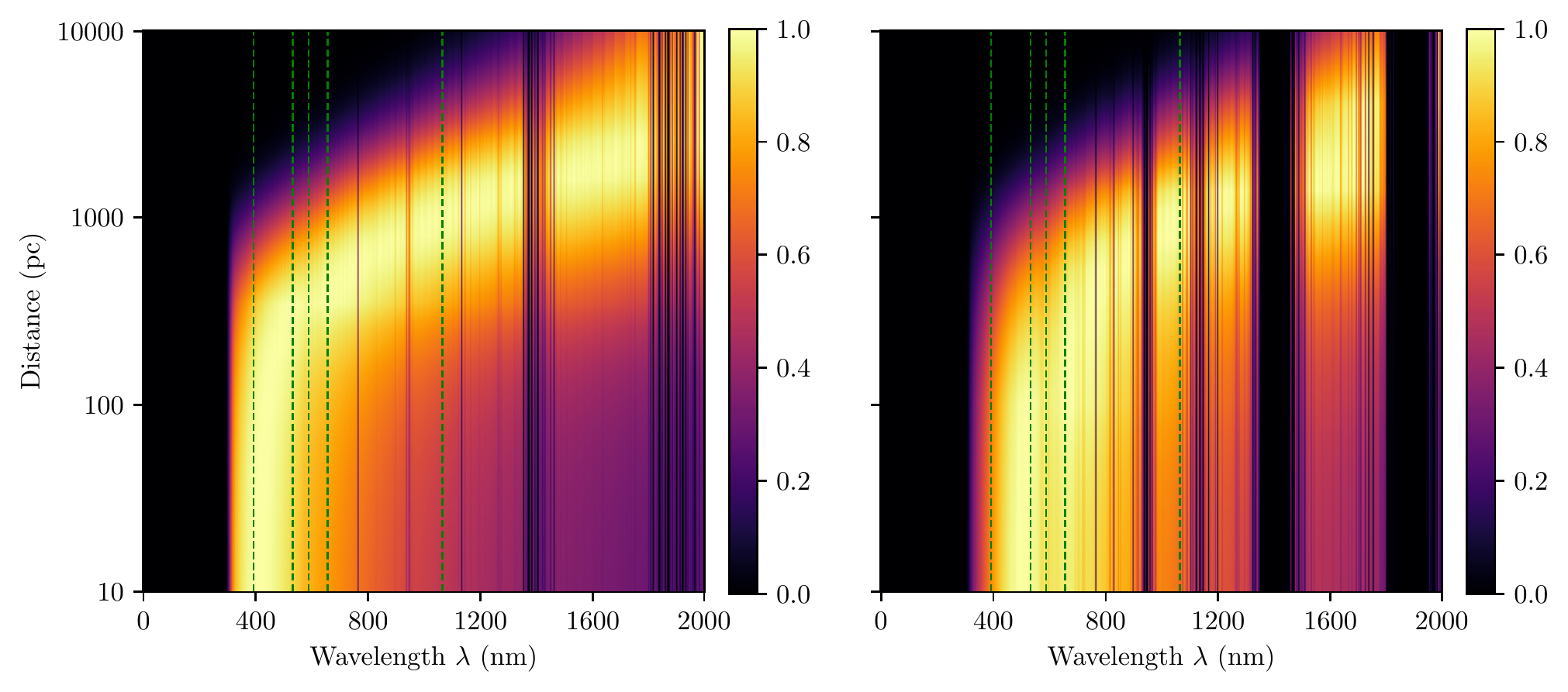}
\caption{\label{fig_photons}Photon throughput as a function of wavelength and distance for the case $S \gg N$. The wavelength-dependent influence of diffraction is included, but has been removed as a function of distance for clarity. Other included factors are extinction and atmospheric transmission. Left: Best atmospheric conditions with 0.5\,mm water vapor at zenith angle. The optical laser lines are optimal within a factor of two out to $\lesssim600\,$pc (393.8\,nm) or $\lesssim\,$kpc (656.5\,nm). For $1<d<3$\,kc, Nd:YAG is optimal within 50\,\%. Right: Lower quartile atmospheric conditions at Cerro Paranal (VLT site) with 5\,mm precipitable water vapor at $20^{\circ}$ zenith angle. The only relevant change is that the shortest laser line (393.8\,nm) moves to the edge of usability at $\sim50\,$\% of maximum flux (532.1\,nm).}
\end{figure*}

\begin{figure*}
\includegraphics[width=.5\linewidth]{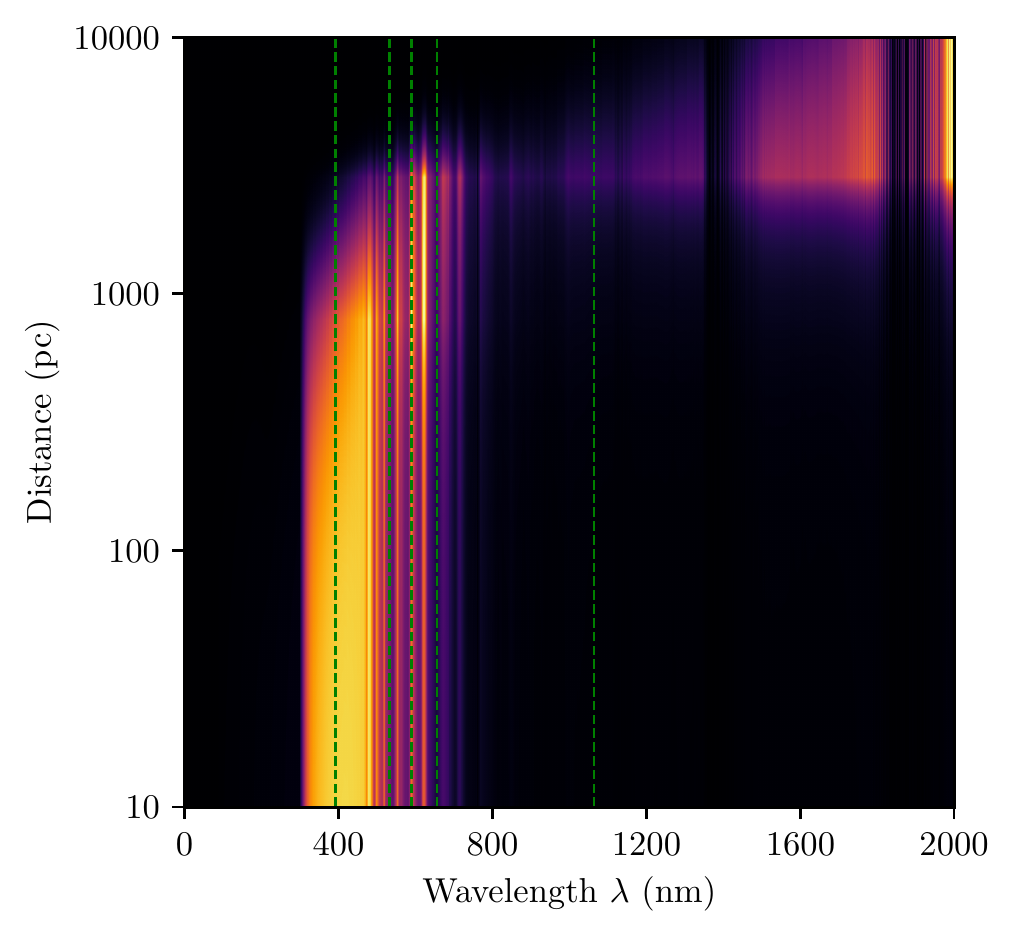}
\includegraphics[width=.5\linewidth]{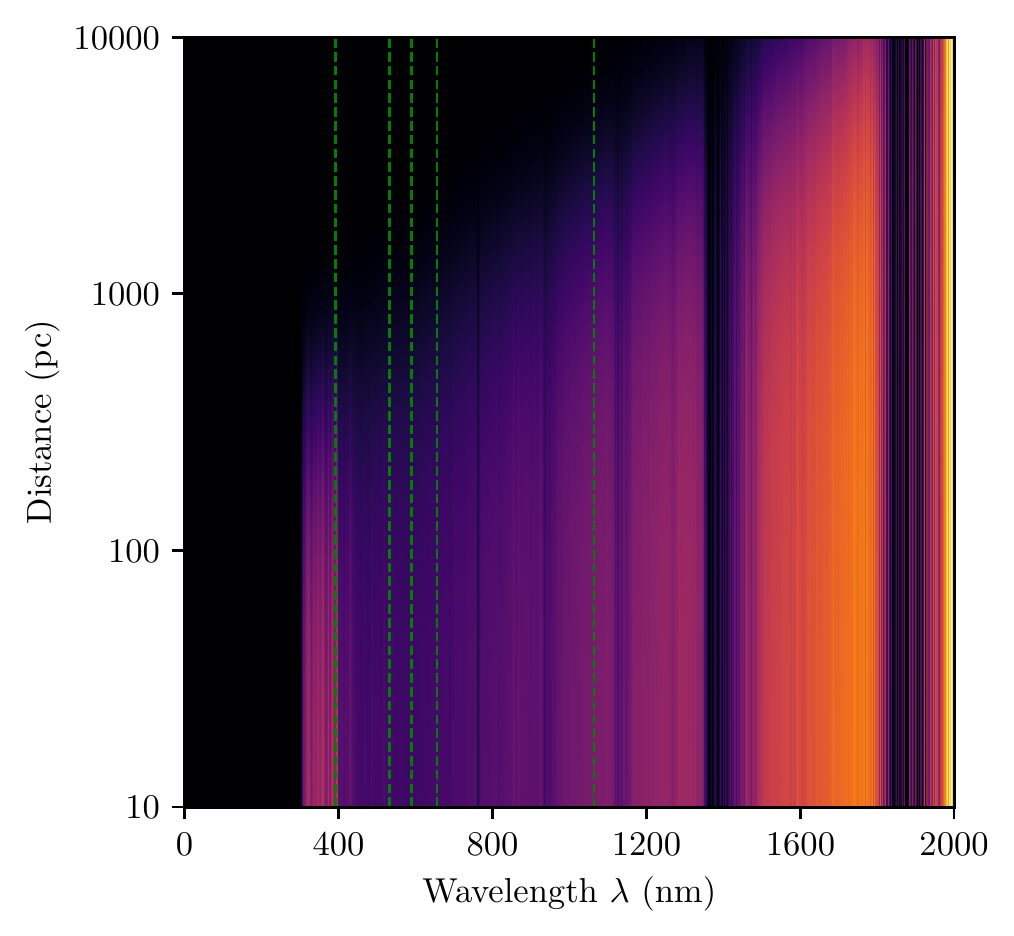}

\includegraphics[width=.5\linewidth]{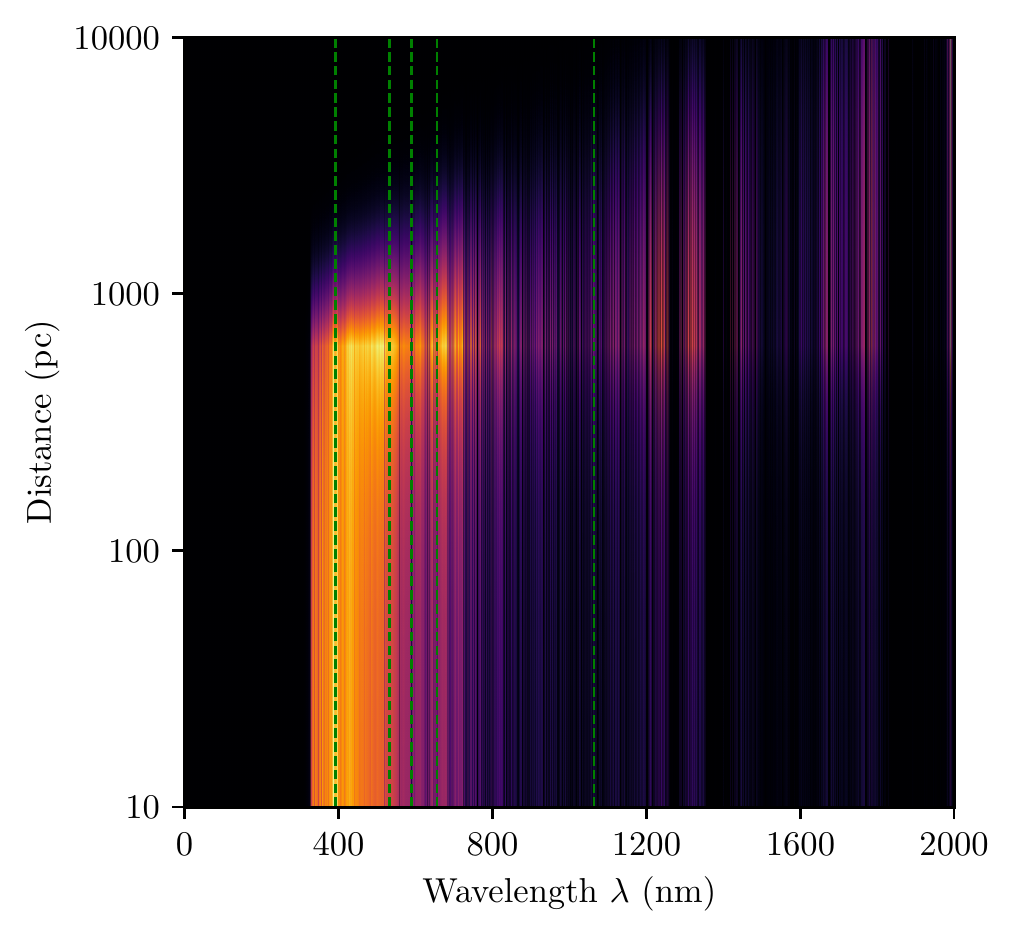}
\includegraphics[width=.5\linewidth]{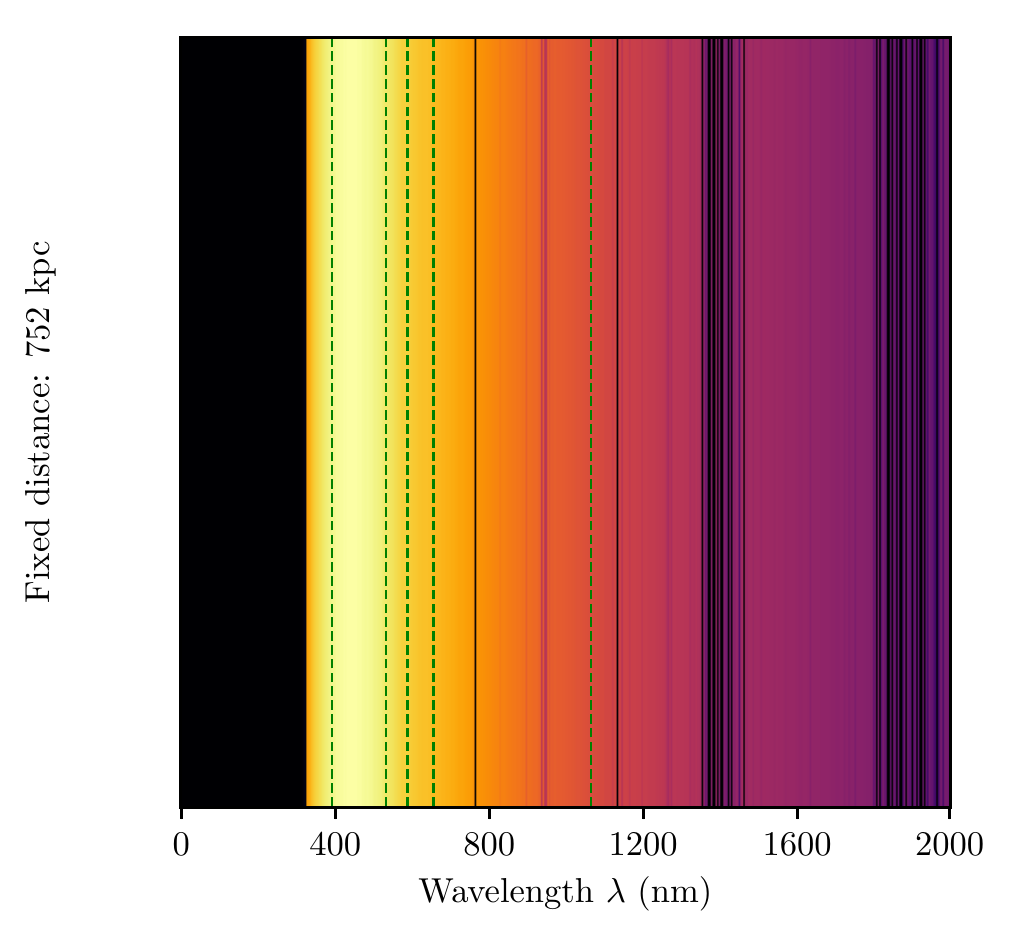}
\caption{\label{fig_photonsNoise}As Figure~\ref{fig_photons}, but including noise at the level $S \sim N$. Top left: M6Ve star Proxima Cen as the background source. Top right: G2V (our Sun). Bottom left: Sky-integrated scattered starlight. Bottom right: Spiral galaxy Andromeda.}
\end{figure*}

\section{Results}
\label{sec:results}
For each wavelength, we can calculate the received flux by assuming some transmitted flux multiplied with the relevant losses from diffraction, extinction, absorption, scintillation, and detector efficiency. The common OSETI assumption is that the signal will be larger than the noise by many orders of magnitude \citep[e.g.,][]{2004ApJ...613.1270H}. In that case, the noise treatment (section~\ref{sub:result_noise}) is irrelevant. 

Laser power as a function of money, however, is relevant. As explained in the introduction, a finite monetary investment has to be considered together with the throughput analyzed in this section. In many cases, the Nd:YAG laser line is within 50\,\% of the most efficient choice, measured as the number of photons received as a function of wavelength. If Nd:YAG lasers produce more than $2\times$ the number of photons compared to other wavelengths, at a given price, they are optimal.

The assumption that $S \gg N$ might be wrong, and the single pulse signal might not be as strong as we believe. Instead, one might imagine repeating weaker pulses, which require a periodicity search. A good example why this is an option comes from lunar laser ranging, which is our closest equivalent to interstellar laser messaging. Lunar laser ranging aims to determine the distance between the Earth and the moon by measuring the light travel time from Earth-based lasers bounced back by retro-reflectors placed on the lunar surface \citep{2008PASP..120...20M,2012CQGra..29r4005M}. Historically, it was common to use few high energy pulses  \citep{1985ITGRS..23..385S,1998A&AS..130..235S,2013RPPh...76g6901M}. Recently, technology has moved to lower pulse energies, but at much higher repetition frequency (80\,MHz) Nd:YAGs at $\lambda=1064\,$nm and $\lambda=532\,$nm with typical widths $\Delta \lambda\sim0.19\,$nm in combination with short (10\,ps) pulse durations \citep{2017CQGra..34x5008A}. This concept is technologically less demanding, because the peak energy in the laser is much lower. Thus, the same money can buy a laser with higher \textit{average} power which emits weaker, but many more, pulses. Only a small fraction of the transmitted pulses is detected, but can be fed into a periodogram search for superior overall accuracy. In this scenario, the wavelength with the highest $S/N$ ratio (plus the monetary argument) is optimal. This concept has been proposed for OSETI by \citet{2013AsBio..13..521L}. It has the additional advantage (over few strong pulses) that the information content (in bits) grows linearly with the number of received pulses.

\subsection{Strong signals}
\label{sub:strong_signals}
Diffraction and extinction are strong functions of distance and wavelength. Absorption is mainly affected by wavelength. Scintillation can be neglected as its effect is small ($\lesssim20\,$)\%, might be positive or negative in a single ns cadence, and is zero on average. The detector efficiency can be assumed as a constant e.g., 50\,\%, and dark counts are assumed to be negligible.

For the signal flux at the receiver, we iterate over the parameter space for good and bad observing conditions and calculate for each
$(\lambda, d)$: $F_{\rm strong}=F_{\rm r}(\lambda, d) \times S_{\rm E} (\lambda, d) \times  S_{\rm A} (\lambda)$. With optimal atmospheric conditions (Figure~\ref{fig_photons}, left panel), optical wavelengths are optimal out to $\lesssim600\,$pc (393.8\,nm) and to $\lesssim\,$kpc (656.5\,nm). For distances $1<d<3$\,kc, Nd:YAG becomes optimal. When atmospheric conditions deteriorate, the shortest optical wavelength end (393.8\,nm) begins to become less attractive. Then, Nd:YAG is the optimal solution, within a factor of two, for all distances $\lesssim 3\,$kpc. The prominent turning point arises from the critical distance where extinction becomes relevant as a function of wavelength.

The lower quartile conditions shown in the right panel might be bad for the VLT site, but still ``great'' at lower altitudes, which represents most of the Earth's surface. As ETI does not know where our observatories are located, it is unclear where the short wavelength cut-off is located; but certainly around $380<\lambda<420\,$nm. Nd:YAG at $\lambda=1{,}064\,$nm is within 50\,\% of the maximum received flux for all conditions and distances.

\begin{figure}
\includegraphics[width=\linewidth]{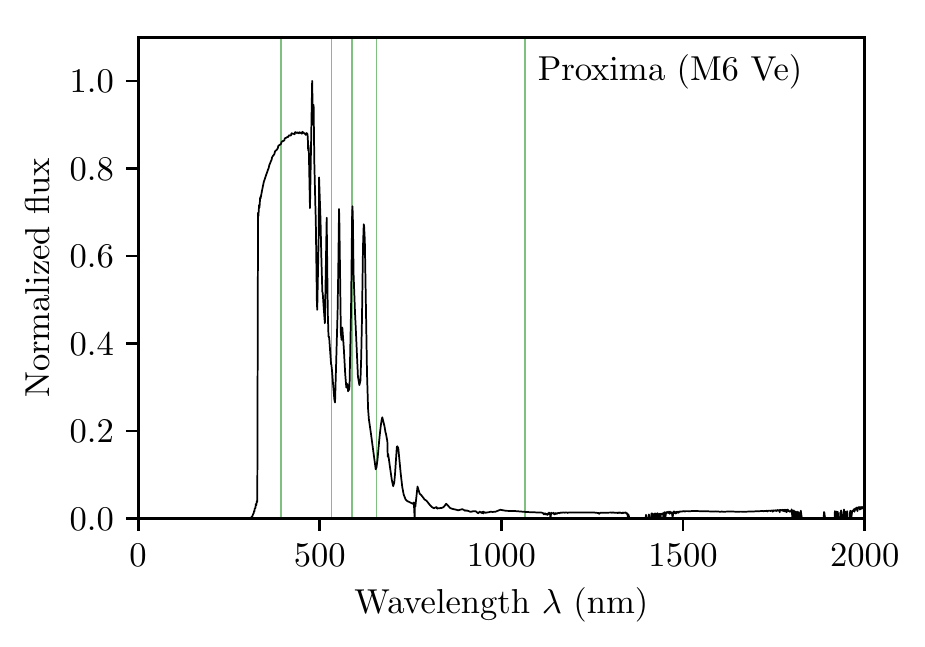}

\includegraphics[width=\linewidth]{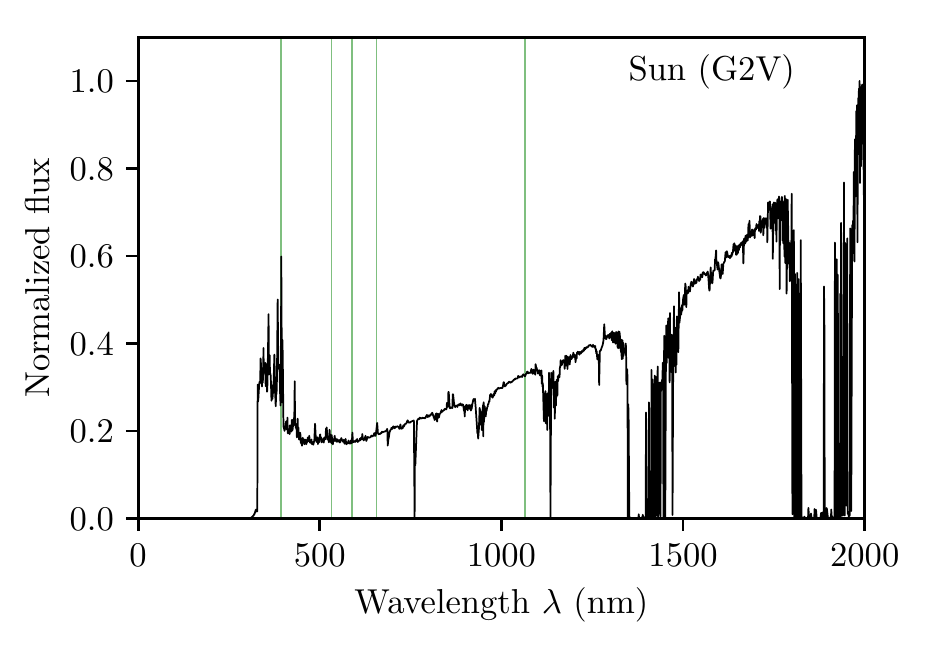}

\includegraphics[width=\linewidth]{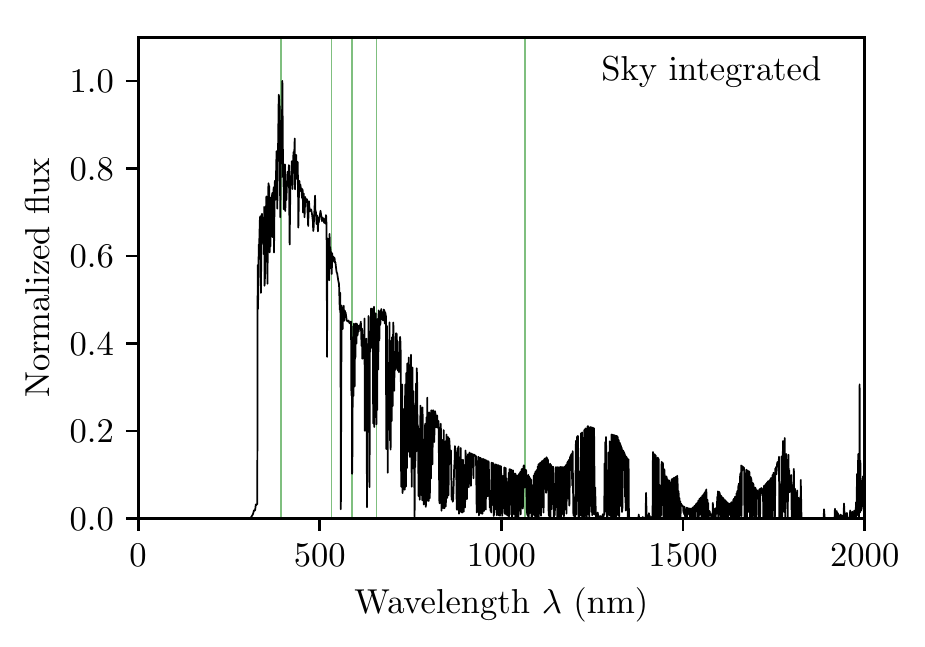}
\caption{\label{fig_slices}Slice through Figure~\ref{fig_photonsNoise} ($S/N$ limited case) for $d=100\,$pc.}
\end{figure}

\subsection{Weak signals}
\label{sub:result_noise}
For weaker signals, we calculate the atmospheric and specific noise (as discussed below) as a function of wavelength: $N=N_{\rm A}(\lambda) + N_{\rm S}(\lambda)$. We take the signal flux from section~\ref{sub:strong_signals} and calculate for each $(\lambda, d)$: $F_{\rm weak}=F_{\rm strong}/N$.

The metric of maximizing $S/N$ breaks down in cases where $S$ is so low that no detection can be made. In our case, atmospheric transparency can be low for some wavelengths, with \textit{very} low noise levels, resulting in high $S/N$. In practice, one would need to establish a minimum flux level given assumptions on receiver and transmitter aperture, and power. To simplify matters, we exclude signal levels which are below $20\,$\% of the maximum. The following Figures are visually similar for cutoff values between $10\dots50\,$\%.

For laser linewidths $\Delta \lambda>1\,$nm, or narrow laser lines which do not coincide with an atmospheric noise peak, we can assume atmospheric noise levels as constant within a factor of two. Atmospheric noise is typically very small ($<10^{-3}$) compared to the flux of even weak single stars. We choose best atmospheric conditions for these examples.

\subsubsection{Individual stars}
The stellar background, when observing one star at a time, is a strong function of the stellar type. M-type stars are preferably observed at the short wavelength end. Also, there are no relevant absorption lines at M-dwarfs near Nd:YAG \citep{2010A&A...523A..58W}. We show the results for a M6Ve star (Proxima Cen) as the background source in Figure~\ref{fig_photonsNoise} (top left) and as a slice for $d=100\,$pc in Figure~\ref{fig_slices} (top). The noise level is set as $S \sim N$, so that the stacking of many pulses is required for a highly significant detection. M-type stars are optimally observed at $320<\lambda<485\,$nm out to kpc. This has been suspected, although without quantification, by \citet{Ross1965}.

Due to their spectral peak near $\lambda\sim\mu$m, Nd:YAG is never useful in the $S/N$-limited case for M-type stars. Distances $d>\,$kpc require $\lambda\sim2\,\mu$m.

Planets around M-type stars might turn out not to be habitable in the vast majority of cases. The habitability of exoplanets around M-dwarfs is heavily debated \citep{2007AsBio...7...30T,2016PhR...663....1S,2017ApJ...845....5K,2017ApJ...846L..21L}. Some of the identified issues include extreme water loss \citep{2015AsBio..15..119L}, volatile deficiency \citep{2007ApJ...660L.149L}, obliquity \citep{2016ApJ...823L..20W}, H/He envelopes \citep{2016MNRAS.459.4088O}, planetary \citep{2011OLEB...41..533L} and stellar magnetic fields  \citep{2013A&A...557A..67V}, and stellar flares \citep{2016ApJ...829L..31D,2017A&A...606A..49P,2018ApJ...855L...2M}. If one or more of these issues turn out to disfavour life around M-dwarfs \citep{2018arXiv180307570L}, these can be excluded from future searches

An important finding for Sun-like stars (G2V, Figure~\ref{fig_photonsNoise}, top right and Figure~\ref{fig_slices}, middle) is that Fraunhofer stellar spectral lines are largely irrelevant, as they are overpowered by the influences of atmospheric transparency and noise, extinction, and diffraction. An exception is perhaps the CaK line (393.4\,nm), if a competitive laser can be made at that wavelength (section~\ref{sub:cak}). For Sun-like stars, longer wavelengths ($\lambda\sim2\,\mu$m) are optimal even for short distances, because their decreased background flux has greater influence than diffraction and atmospheric noise. Optical and NIR wavelengths are inferior by a factor of a few in the $S/N$ limited case.

\subsubsection{Sky-integrated surveys}
Surveys with large fields of view collect mostly atmospheric noise and scattered starlight. The optical flux is lower by a factor of a few compared to IR, so that the shortest laser line (CaK) is optimal out to $\sim\,$kpc. Nd:YAG in the IR is worse by an order of magnitude (Figure~\ref{fig_photonsNoise}, bottom left and Figure~\ref{fig_slices}, bottom). An all-sky survey for weak sources within a sphere of radius kpc should strongly favor the Fraunhofer CaK line at 393.4\,nm with a narrow filter (again, if a competitive laser can be made at that wavelength).

\subsubsection{Andromeda galaxy}
As another use case, we test for the optimal laser line from the Andromeda galaxy. At a distance of $752\pm27$\,kpc \citep{2012ApJ...745..156R}, diffraction is the strongest factor. A 10\,m (km) transmitter at $\lambda=\,\mu$m located in Andromeda illuminates a disk with a size of 0.1\,pc (200\,au) in our galaxy. Thus, it is still aimed at a single stellar system. Foreground extinction (galactic and intergalactic) towards Andromeda is surprisingly low, A(V)=0.17 \citep{2011ApJ...737..103S}, comparable to a distance through the galactic disk of 100\,pc, so that $S_{\rm E}>0.75$ for all spectral lines.

With a brightness of $\sim 4\,$mag for the Andromeda galaxy, its flux is $6\times10^8\,\gamma\,$s$\,^{-1}$m$^{-2}$. A laser of equal brightness would require very high power. Following Eq.~\ref{eq2} and setting $\lambda=532\,$nm, $D_{\rm t}=1\,$km, it requires $P=10^{17}\,$W, or $10^6\times$ more than the proposed ``Starshot'' beamer (100\,GW).

Using the same $D_{\rm t}=D_{\rm r}=10\,$m telescopes as in the introduction, a pulsed laser would require a pulse energy of $10^{5}\,$MJ, or $10^5\times$ more than the most powerful laser on Earth, to detect on average one photon per pulse. While a MJ pulse compressed to ns duration has a peak energy of $10^{15}\,W$, the bottleneck is diffraction and flux. In both cases, the receiver simply does not collect sufficiently many photons. These scenarios are already built upon very tight beams, illuminating only one stellar system at a time. For broader beams, in the extreme case as a beacon for our whole galaxy, the energy requirements would be larger by many orders of magnitude.

Despite these high requirements, we show the optimal wavelengths in Figure~\ref{fig_photonsNoise} (bottom right). Optical wavelengths are preferred (e.g., SHG-Nd:YAG at 532\,nm), but NIR Nd:YAG is within a factor of two.

\section{Discussion and conclusion}

\subsection{Previous and future searches}
\label{sub:previous_searches}
After the seminal OSETI paper by \citet{1961Natur.190..205S}, searches for pulsed signals were performed in the MANIA project \citep{1993ASPC...47..381S} and at Columbus Observatory \citep{1993SPIE.1867..178K,1995ASPC...74..387K}. Observations with a time resolution of $3.3\,\mu$s were performed by \citet{1997Ap&SS.252...51B}, who also discuss spectral versus temporal coding. An extensive survey was started with the Harvard-Princeton search \citep{2004ApJ...613.1270H} covering $6{,}000$ objects during $2{,}400$ observing hours. This effort was extended to all of the sky \citep{2007AcAau..61...78H} using a drift scan. Other projects include e.g., Berkeley's search with the $0.8\,$m automated telescope at Leuschner observatory \citep{2011SPIE.8152E..12K}, and observations at Lick Observatory \citep{2004IAUS..213..415W,2005AsBio...5..604S}. While most surveys used classical telescopes, there have also been efforts using Cherenkov-based instrument \citep{1996APh.....5..353O,2001AsBio...1..489E,2005ICRC....5..387H,2009AsBio...9..345H}. There are few searches in the infrared \citep[NIROSET,][]{2014SPIE.9147E..0JW,2014SPIE.9147E..4KM,2016SPIE.9908E..10M}. In addition to surveys, individual objects have been examined in detail, such as the exoplanet host stars Trappist-1, GJ 422, Wolf 1061 \citep{2018AAS...23110401W} and the anomalous Boyajian's Star \citep{2016ApJ...825L...5S,2016ApJ...818L..33A}. A space mission with a nanosatellite has also been proposed \citep{kayal2017nanosatellite}.

In the regime of continuous narrowband signals, spectral searches have been performed \citep{2002PASP..114..416R,2015PASP..127..540T,2017AJ....153..251T}.
A more exotic method uses spectral modulation of coherently separated laser pulses, as first noted by \citet{1992ApOpt..31.3383C} and subsequently studied in \citet{2010ApJ...715..589B,2010A&A...511L...6B,2012AJ....144..181B} with the motivation to apply the method to astronomical data. Searches for such periodic temporal modulations were reported for SDSS spectra of galaxies \citep{2013ApJ...774..142B}  and stars \citep{2016PASP..128k4201B,2017JApA...38...23B}.

With the learnings from this paper, future searches can employ narrow line filters (e.g., 1\,nm) centered at e.g., $532.1\,$nm. This allows for large fields of view (a significant fraction of the sky) and fast survey speeds. A second scenario is discussed in the following section.

\begin{figure}
\includegraphics[width=\linewidth]{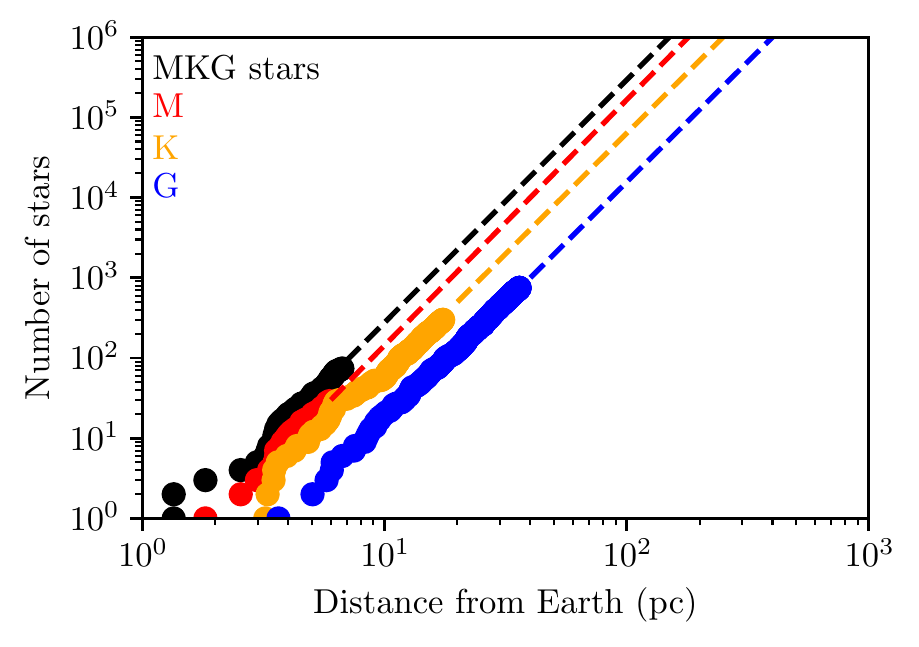}
\caption{\label{fig:hip}Number of stars within a spherical volume as a function of distance. There are $10^6$ G-dwarfs within 300\,pc. Data from the Hipparcos catalog \citep{1997ESASP1200.....E,1997A&A...323L..49P}.}
\end{figure}

\subsection{What distance should we look at?}
As discussed in section~\ref{sec:results}, the optimal wavelength moves from optical to IR for distances $d>\,$kpc. How many civilizations inside a sphere of one kpc are signaling towards us? If it is less than one, we should search in the infrared.

We do not know of any other life outside of Earth, so we do not know their distances. We can however estimate the distances to potentially habitable exoplanets. The fraction of stars that have planets is of order unity \citep{2014ApJ...791...10M}. The fraction of stars with rocky planets in the habitable zone ($\eta{\rm -earth}$) has been estimated between $2\pm1$\,\% \citep{2014ApJ...795...64F} and $11\pm4$\,\% \citep{2013PNAS..11019273P} for sun-like stars (G- and K-dwarfs) and 20\,\% for M-dwarfs \citep{2015ApJ...807...45D}. The habitability of planets around the more numerous M-dwarfs is unclear \citep{2016arXiv161005765S}. Also, the fraction of planets in the habitable zone which actually develop any sort of life ($\eta{\rm -life}$) is unknown. Using the parameters in the  \citet{2013IJAsB..12..173D} equation, one could estimate some fraction of stars $\eta_{\rm i}$ with intelligent life signaling in the optical or IR towards us.

In Figure~\ref{fig:hip}, we show the number of stars by spectral type as a function of distance from the Earth. For example, there are $10^6$ G-dwarfs within 300\,pc. If only one of these is signaling towards us, the expected average distance is $1 / \sqrt{2} d_{\rm G} \sim 212$\,pc ($\sim90\,$pc for all MKG-stars).

This estimate includes stars of any age, while stars in the inner (thin) disk are generally younger, $0 \dots 8$\,Gyr with a median age of $\approx4$\,Gyr, and stars in the thick disk are older \citep[$8 \dots 12$\,Gyr,][] {2013A&A...560A.109H}. For reference, the total number of stars in the galaxy is $\approx2\times10^{11}$.

This order-of-magnitude estimate holds for the width of the galactic disk \citep[600\,pc,][]{2013A&ARv..21...61R}, corresponding to $3\times10^{6}$ G-stars ($10^{8}$ MKG-stars). There is likely a ``galactic habitable zone'' (GHZ), argued to range from 4 to 10 kpc from the center of the galaxy (the sun is at 8\,kpc), with 10\% of galactic stars in the GHZ, and 75\% of these stars in the GHZ are older than the sun \citep{2004Sci...303...59L}. Other estimates see 1\% of galactic stars in the GHZ \citep{2011AsBio..11..855G}.

It has been argued that advanced civilizations colonize and inhabit ``clusters'' in space \citep{2016AsBio..16..418L}. The number of clusters would mainly depend on their numbers and longevity. For lifetimes $>1$\,Myr, a single club can be established; for shorter lifetimes, several groups might emerge \citep{2016arXiv160808770F,2017arXiv170703730F}.

To conclude, if the fraction of ``signaling'' stars is $\eta_{\rm i}>10^{-7}$ (for G-type stars) or $\eta_{\rm i}>10^{-8}$ (for MKG stars), then we should search in the optical within 1\,kpc. Otherwise, larger distances and IR wavelengths are favored.

\begin{figure}
\includegraphics[width=\linewidth]{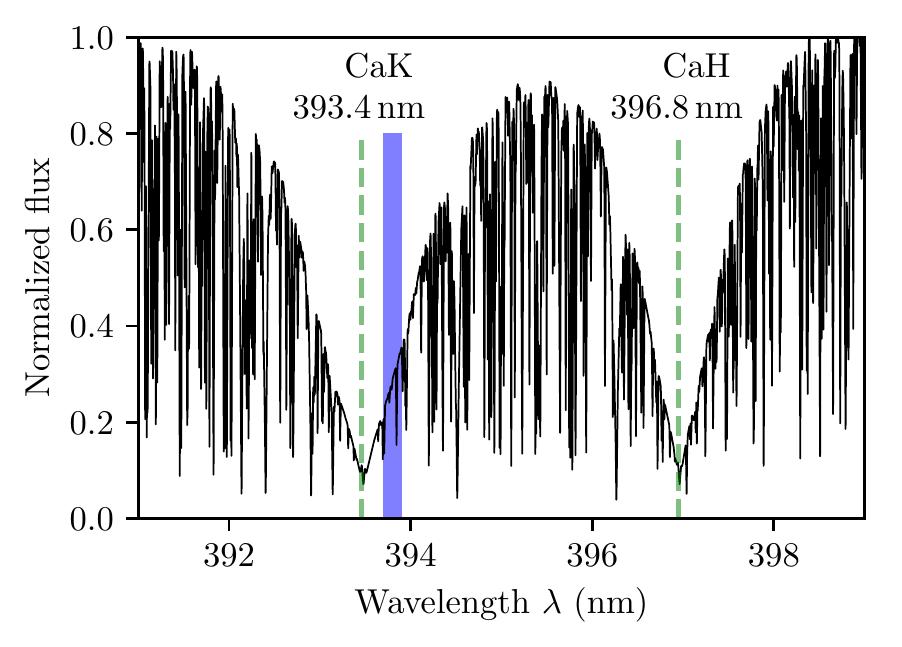}
\caption{\label{fig:figure_oseti_sun_spectrum_zoom}High resolution solar spectrum \citep{2005MSAIS...8..189K} near CaK and CaH (green lines). The laser line as discussed in the text is shown with $\Delta\lambda=0.2\,$nm (blue shade).}
\end{figure}

\subsection{Fraunhofer lines versus laser lines}
\label{sub:cak}
The secondary laser lines, as discussed by \citet{2018NewA...60...61N}, are applicable in case the primary Nd:YAG choice is strongly disfavored. For example, a wavelength of 393.8\,nm close to the CaK line can be produced with a sum frequency generation of YAG(R$_2\rightarrow$X$_3$) plus the second harmonic of YAG(R$_1\rightarrow$Y$_5$) (Table~\ref{tab:laser_lines}). The authors have not judged the power efficiency of this laser type, so this remains to be validated by laser engineers. 

As can be seen in Figure~\ref{fig:figure_oseti_sun_spectrum_zoom}, this exemplary laser is close, but not exactly at, the CaK line. Depending on the construction of a laser, it is possible to tune it by a few 0.1\,nm. The difference of 0.4\,nm corresponds to a velocity of 305\,km\,s$^{-1}$, which might partly be compensated by the proper motions between the transmitter and Earth.

The linewidth shown in Figure~\ref{fig:figure_oseti_sun_spectrum_zoom} is $\Delta \lambda = 0.2\,$nm. Lasers with broader linewidth would average the flux near the Fraunhofer line, making it less attractive. As discussed in paper 11 of this series, the optimal temporal pulse width of OSETI pulses is $\approx1\,$ps which limits $\Delta \lambda \gtrsim 0.5\,$nm for $\lambda_0=393.4\,$nm through Heisenberg's uncertainty principle.

\subsection{Unknown laser power and technological level}
The strongest lasers on Earth supply $\approx2\,$MJ in a 5\,ns pulse, at low (hrs) repetition rates \citep{2017ApPhB.123...42H}. This technological level provides the use-case for ongoing observations (section~\ref{sub:previous_searches}). Our own technology might allow for much brighter and tighter beams in the near future. A laser-pushed lightsail \citep[''Breakthrough Starshot'',][]{2016arXiv160401356L,2016Sci...352.1040M,2017Natur.542...20P} was suggested for a mission to $\alpha$\,Cen. A km-sized phased aperture would emit 100\,GW of laser power, sufficient to accelerate a 1\,g ``space-chip'' to $v=0.2\,$c in minutes. If this beamer is targeted at the Planet Proxima~b directly, it would appear as a star of magnitude $-4$ on its sky, comparable to Venus and visible in daylight. Even over a distance of 100\,pc, it would produce a flux of $10^{10}$ photons m$^{-2}$\,s$^{-1}$, resembling a bright first magnitude naked eye star \citep{2016SPIE.9981E..0HL}. The beam falls below the naked eye visibility (6\,mag) at a distance of $\sim2\,$kpc. Astrometry, proper motion estimates and pointing accuracy must be of the same order ($\mu$as) as required for spaceflight, a level which is achievable with future telescopes for nearby stars \citep{2017AJ....154..115H}.

As humans have not, over the recorded history, observed a second Sun in the sky, we can exclude that such beamers are commonly targeted at Earth. More precisely, we can only exclude such signals in the wavelength range of the human eye between 390 and 700\,nm \citep{starr2006biology}. A Nd:YAG laser near $1{,}064\,$nm would only be visible to infrared detectors, which are far less common than optical instruments. Incidentally, ``Starshot'' baselines a Nd:YAG laser for their beamer \citep{Kulkarni2018}. A shallow, all-sky all the time survey for such transients appears to be attractive.

\acknowledgments
\textit{Acknowledgments}
MH is thankful to Marlin (Ben) Schuetz for useful discussions.

\clearpage
\pagebreak

\end{document}